\documentclass[11pt,a4paper]{article}
\usepackage[utf8]{inputenc}
\usepackage{authblk}
\usepackage{graphicx}			
\usepackage{amssymb}
\usepackage{amsmath}
\usepackage{cite,url}

\newcommand{\be}{\begin{equation}}
\newcommand{\ee}{\end{equation}}
\newcommand{\bea}{\begin{eqnarray}}
\newcommand{\eea}{\end{eqnarray}}

\newcommand{\D}{\mathcal{D}(\epsilon)}
\newcommand{\e}{\epsilon}

\title{{\bf Compact stars in quantum spacetime}}
\author[1]{E. Harikumar\footnote{harisp@uohyd.ernet.in} and Zuhair N S \footnote{zuhairns@gmail.com}}
\affil[1]{\small{School of Physics, University of Hyderabad, Central University P.O.,
Hyderabad 500046, India}}
\date{}
\begin{document}
\maketitle
  \begin{abstract}
We derive bounds on the deformation parameter of the $\kappa$-spacetime by analyzing the effect of non-commutativity on astrophysical model. We study compact stars, taken to be degenerate Fermi gas, in non-commutative spacetime. Using tools of statistical mechanics, we derive the degeneracy pressure of the compact star in $\kappa$-spacetime and from the hydrostatic equilibrium conditions we obtain a bound on the deformation parameter. We independently derive this bound using generalized uncertainty principle, which is a characteristic feature of quantum gravity approaches, strengthening the bound obtained.
  \end{abstract}

\section{Introduction}

Formulating theory of gravitation in the quantum framework is quintessential for a complete understanding of the laws of universe. Recently, there has been attempts along different directions to explain quantum gravity. All these quantum gravity models predict the existence of a minimal length scale, at which the features of quantum gravity becomes discernible. One such feature is that the concept of spacetime becomes meaningless in an operational sense \cite{Doplicher:1994tu}. Implementing non-commutativity in spacetime is an efficient approach to capture the structure of spacetime at Planck scale\cite{Doplicher:1994tu}. 

One of the physical systems, where quantum gravitational aspects are relevant, is in the situation of enormous amount of particles packed into a compact region of spacetime. As the region becomes more and more compact, we find that gravitational strength become so enormous that we have to take into account the features of quantum gravity theory. Since statistical mechanics plays an important role in understanding a system consisting of large number of particles, it is reasonable to assume that a study of statistical mechanics of these compact system in the background of a noncommutative spacetime will be a viable option to understand the Planck scale physics. The effect of Planck scale physics, especially of non-commutativity, in statistical mechanics has been reported by many in the literature\cite{Khan:2007fc,KowalskiGlikman:2001ct,AmelinoCamelia:2009tv, Shariati:2010zz,Scholtz:2012tx,Hosseinzadeh:2015jra}. Compact stars appears to be a potential source to observe the effects of non-commutativity on statistical mechanics, due to its accessibility for observation. In this letter, we will be focusing primarily on the compacts stars to study the effect of non-commutativity.

There is another reason for the choice of compact stars as a means to explore the effects of noncommutativity. Astrophysical phenoemena have been attracting attention in the context of searches for quantum gravity signatures in recent times. Recently, there have been many interesting works searching for pointers of quantum gravity in gamma ray bursts\cite{AmelinoCamelia:1999pm, AmelinoCamelia:2008qg, Borowiec:2009ty}. The presence of GZK limit is studied in the literature as a manifestation of Planck scale physics\cite{AmelinoCamelia:2001vz}. Another interesting application of non-commutative physics is used in the study of GRB time delay\cite{Li:2009mt}. It was argued that observations on Ultra High Energy Cosmic Rays (UHECR) are capable of providing indication about the modifications due to non-commutative physics on Hanbury-Brown Twiss (HBT)\cite{Srivastava:2012av}. However, in the present paper, we take compact stars as an example to look for effects of quantum gravity at the astrophysical scale. Compact stars is one of the familiar example, where quantum mechanics and gravity comes in to play together. 

The non-commutativity of space-time introduces a natural length scale beyond which sensible measurement of spacetime becomes meaningless. The existence of this length scale introduces an uncertainty in the measurement of spacetime coordinates. This led to the possibility of modification of Heisenberg uncertainty principle to a generalised uncertainty principle(GUP) \cite{Amati:1988tn, Garay:1994en,Kempf:1996nk,AmelinoCamelia:1997jx}. The assumption of length scale leads to a deformation of Heisenberg algebra. It is shown that this deformation depends explicitly on a dimensionful parameter of mass dimension\cite{Maggiore:1993kv}. Further, this deformed algebra is shown to establish the underlying algebraic structure for a GUP relation\cite{Maggiore:1993kv}. The implications of generalised form of uncertainty principle have been a source of investigation for the past few decades. A possible way to recover the uncertainty principle is to study the quantum aspects of geometry\cite{Capozziello:1999wx}. GUP is a useful tool to understand the corrections to Hamiltonian at the quantum gravity scale. Physical models have been studied using GUP and the corresponding quantum gravity corrections have been calculated \cite{Das:2008kaa}. These computations are used to obtain an upper bound on the length scale relevant to quantum gravity. An area of interest where the quantum gravity effects becomes prominent and is probably measurable, in the near future, is at the cosmological scale. Deformations to dispersion relation due to GUP is used to study ultra-relativistic particles in the early universe. It was shown that GUP leads to modification of thermodynamic behavior of these particles\cite{Nozari:2006gg}. This suggest the possibility of studying thermodynamics and statistical mechanics of systems at quantum gravity scales using GUP deformations. The existence of a generalised uncertainty principle will also lead to a minimum accessible phase space volume. This has interesting consequences on the density of states of statistical mechanics system. The feature of decrease in the availability of microstates in the quantum gravity regime is explained by invoking generalised uncertainty principle \cite{KalyanaRama:2001xd}. 

The above mentioned feature of reduction of microstates is easily illustrated in the study of degenerate Fermions. Since compact stars are an example of degenerate systems and in addition compact stars have the advantage of observational accessibility, it is of interest to study the effect of non-commutativity on the density of states and its effect on physical systems. In this work, we study compact stars, which is made up of degenerate Fermi gas, as a potential platform to study the corrections due to quantum gravity. This is done in two different ways. We first use modifications in tools of statistical mechanics due to non-commutativity to study the compact stars. We then re-derive these results in a completely different approach where GUP is used. Such a study invites the possibility of extending to the analysis of black holes in noncommutative spacetime. For this reason, the study of compact stars can be thought of as a first step in understanding the effects on black holes at the quantum gravity level and thus to have a better grasp on the features of quantum gravity theory. 

In this letter, we use a specific example of noncommutative spacetime, namely $\kappa$-spacetime. $\kappa$-spacetime is the spacetime, one in which it's space and time coordinates are noncommuting and spatial coordinates commutes with each other.
More specifically, it satisfy the relation: $[\hat{x}_0,\hat{x}_i]=ia\hat{x}_i$ and $[\hat{x}_i,\hat{x}_j]=0$. In recent literature, it has been shown that the introduction of spatial noncommutativity lifts the degeneracy in fermion gas\cite{Bemfica:2005pz}. In this letter we will, however, be analyzing the effect of space and time noncommutativity using $\kappa$-spacetime. In this letter, we follow the approach of modelling $\kappa$-deformed theory in terms of commutative variables and deformation paremeter, which is constructed using a realization of $\kappa$-deformed coordinates in terms of commutative coordinates and their derivatives\cite{Meljanac:2006ui}. In this letter, we start with a choice of realization, $\varphi= e^{\mp ap^0}$ in $\kappa$-spacetime. It is to be recalled that non-commutativity replaces usual particle statsitics with a twisted statistics. It is thus possible to study the effect of noncommutativity by applying twisted statistics\cite{Chaichian:2004za, Balachandran:2006pi, Govindarajan:2008qa, Khan:2007fc}. However, in the approach used in this letter, all the effects of $\kappa$-deformation is incorporated through the realization of non-commutative coordinates and the corresponding deformed dispersion relation. By this approach, we could work with ordinary Fermi statistics for Fermi gas in noncommutative spacetime, expressed in terms of commutative coordinates, their derivatives and deformation parameter $a$. With this idea in mind, we use the dispersion relation in this realization to find the density of states for a given energy interval. So obtained density of states is then used to calculate the energy density, number density and pressure due to gas of degenerate Fermi gas.  These expressions are used to obtain the equation of state in the non-commutative case. In our calculation, since we are interested only in the region where density of matter is very high, we restrict ourselves to relativistic regime. As Fermi momentum is related to number density, this indicates the region we are looking at is where the Fermi momentum satisfy, $p_F \gg m_e c$.  We then apply these results to stars modelled as degenerate Fermi gas. Since, the pressure due to degeneracy opposes the gravitational force, the equilibrium situation implies the balance of gravitation pressure with radiation pressure. This is then used to obtain an upper bound on the value of deformation parameter $a$. We then start from the GUP and using it, derive the degeneracy pressure of Fermion gas. Using this we obtain a bound on $a$, which exactly matches with the previously obtained bound.
\section{Degeneracy pressure and hyrdrostatic equillibrium in $\kappa$-spacetime: A statistical mechanics approach}
In this section we use tools of statistical mechanics to derive degeneracy pressure of gas of Fermions in $\kappa$-deformed spacetime. Since the compact stars are modelled as degenerate Fermi gas, this allows us to analyze properties of compact stars in $\kappa$-deformed spacetime. First, we will briefly review the calculations in commutative spacetime. Standard calculation in commutative settings starts with the computation of partition function for a Fermionic gas. The partition function is then used to obtain the formula for pressure, by using the relation between pressure and Landau potential, in terms of energy density. The density of states derived using energy-momentum relation is then used to find the hydrostatic equillibrium conditions. In performing this step, the Fermi gas is assumed to be degenerate, so that the temperature of the gas is safely set as $T=0$. In order to study the balance between degeneracy pressure and gravitational pressure, the pressure obtained above is equated with gravitational pressure to set the condition for the equilibrium. This study of compact stars using statistical mechanics is crucial in understanding the stability of stars under gravitational collapse and for the in-depth exploration of the possibility of gravitational singularity (details on calculations in commutative spacetime can be seen in any standard textbooks on statistical mechanics, see for eg.\cite{huang, pal}) and this led to famous Chandrasekhar limit for white dwarf \cite{Chandrasekhar:1931ih}.

We start with the dispersion relation (see for details \cite{Meljanac:2006ui, Meljanac:2007xb}) in $\kappa$-space time given by
\bea 
\frac{4 k^{2}}{a^{2}} \sinh^{2}(\frac{ap^{0}}{2k}) - p_1^{2} \frac{e^{-\frac{ap^{0}}{k}}}{\varphi^{2}\left(\frac{ap^{0}}{k}\right)} + \frac{a^{2}}{4 k^{2}} \left[ \frac{4 k^{2}}{a^{2}} \sinh^{2}\left(\frac{ap^{0}}{2 k}\right) -  p_1^{2} \frac{e^{-\frac{ap^{0}}{k}}}{\varphi^{2}(\frac{ap^{0}}{k})} \right]^{2} = m^{2}c^{2}. \label{disp}
\eea
Note that in our calculation, we take into the possibility of replacing of $\hbar $ with $k = \hbar + \frac{a^2 c^3}{G}$, which has the correct dimension and reduces to $\hbar$ in the commutative limit $a \rightarrow 0$. The possibility of such a modification was studied in \cite{Scholtz:2012tx}.

\bigskip 

For the choice of realization, $\varphi = e^{\mp \frac{ap_0}{kc}}$, we have the energy-momentum dispersion relation, valid to first order in $a$, is of the form
\be 
p_0^{2}c^{2} - p_i^{2} (1\pm \frac{ ap_0}{k}) c^{2} = m^{2}c^{4}
\ee
In writing the above equation, it should be emphasized that the factor $k$ contains term having second order in $a$, but we retain only the first order term in the binomial expansion of $k$.

This allows two possible solutions for energy $p_0$ as
\be 
p_0 = \frac{1}{2} \left( \pm \frac{a}{k} p_i^2-\sqrt{\left(\frac{\pm a}{k}\right)^2 p_i^4+4 m^2c^{2}+4 p_i^2}\right), \, \, p_0 = \frac{1}{2} \left(\sqrt{\left(\frac{\pm a}{k}\right)^2 p_i^4+4 m^2c^{2}+4 p_i^2}\pm \frac{ a}{k} p_i^2\right).
\ee
Among these, the solution with correct commutative limit is
\be 
p_0 = \frac{1}{2} \left(\sqrt{\left(\frac{ \pm a}{k}\right)^2 p_i^4+4 m^2c^{2}+4 p_i^2}\pm \frac{a}{k} p_i^2\right).
\ee
It is to be noted that the `$\pm$' has its origin from two different possible choice of $\varphi$ realizations. Keeping leading order correction in $a$, we write
\be 
p_0 =  \left(\sqrt{m^2c^{2}+p_i^2} \pm \frac{1}{2}\frac{ a}{k} p_i^2\right) .
\ee

We denote the non-commutative energy by $\e$ and commutative energy by $\tilde{\e}= \sqrt{p^{2}c^{2}+m^{2}c^{4}}$, so above equation takes the form
\be 
\e = \sqrt{m^2c^{4}+p_i^2c^{2}} \pm \frac{1}{2}\frac{ a}{k} p_i^2c = \tilde{\e} \pm \frac{1}{2}\frac{ a}{k} p_i^2c.
\ee
Taking differentials, we obtain (keeping terms to first order in $a$) 
\bea 
d\e &=& d\tilde{\e} \pm \frac{ ac}{k}pdp \label{detile}
\eea
From now on, we will denote the magnitude of $\vec{p}$ as $p$. Using the expression for commutative energy, we find
\be 
\tilde{\e} d\tilde{\e} = c^{2} p dp \label{tilep}
\ee
Now, using eqn. \eqref{tilep}, derivative of $\e$ with respect to $p$ is,
\be 
\frac{d\e}{dp}= \frac{d\tilde{\e}}{dp} \pm 2 \frac{ ac}{k}p =  \frac{pc^{2}}{\tilde{\e}} \pm  \frac{ac}{k}p. \label{ncdedp}
\ee
 Working in the non-commutative canonical phase space $(\{x\},\{p\})$, we write the density of states per unit volume per unit energy for an electron gas as

\be 
\mathcal{D(\tilde{\e})} = \frac{4\pi g p^2}{(2\pi k)^3}\frac{dp}{d\tilde{\e}}. \label{ncdensity1}
\ee
Note that $\mathcal{D(\tilde{\e})}$ become $a$ dependent through $k$ and $\frac{dp}{d\tilde{\e}}$. This dependence on $k$ leads to new corrections which are first order in $a$.

From eqns.\eqref{ncdensity1} and \eqref{ncdedp}, we write
\be 
\mathcal{D(\tilde{\e})} = \frac{4\pi g \vec{p}^2}{(2\pi k)^3} \frac{\tilde{\e}}{pc^{2}} \frac{1}{(1\pm \frac{ a\tilde{\e}}{ck})}= \frac{4\pi g \vec{p}}{(2\pi k)^3} \frac{\tilde{\e}}{c^{2}} (1\mp \frac{ a\tilde{\e}}{ck})
\label{ncdensity2}
\ee
where, $g=2$ for an electron. Recall that the presence of $\pm$ in the expression is due to the reason that we have two different choices of $\varphi$ realizations permissible. We thus see that presence of non-commutativity leads to a modification in the density of states, which is proportional to the deformation parameter $a$ and energy of the state $\tilde{\e}$. Also, note that in the limit $a\rightarrow 0$, we find that the expression for density of states matches with the familiar commutative result.

\bigskip

Probability for a system to be in state j with energy $\e_j$, particle number $N_j$ and  chemical potential $\mu$ is given by the well known relation
\be 
\mathcal{P}_j = \frac{1}{Z} e^{\beta (\mu N_j -\e_j)}
\ee
with $\beta = 1/(k_BT)$. Then, $Z$ is given as
\be 
Z = \prod_r \sum_{n_r} \exp\left( - \beta \left( \e_{n_r} - \mu N_{n_r} \right) \right),
\ee 
with $n_r$ being the number of particles in the state labelled $r$.

The average number of particles in state $r$ is
\be 
\bar{N_r} = \frac{1}{Z} \sum_n N_r \exp\left( - \beta \left( \e_{n_r} - \mu N_{n_r} \right) \right).
\ee

Since electrons obey Pauli's exclusion principle, 
\be 
Z = \prod_r \sum_{n_r=0,1} \exp\left( - \beta \left( \e_{n_r} - \mu N_{n_r} \right) \right)
\ee
This implies,
\be 
\ln Z = \sum_{all \, states \, r} \ln \left( 1+e^{-\beta(\e_r-\mu)} \right). \label{lnz}
\ee
 It is to be stressed that in the expression for partition function, we have the effect of $\kappa$-deformation coming through the correction to the energy.

Recall that the Landau potential is connected to $Z$ as,
\be 
\Phi = - k_B T \ln Z. \label{phitoz}
\ee
Hence, the Landau potential $\Phi$ is given by
\be 
\Phi(T,V,\mu) = U-TS- \mu N = - pV.
\ee
We get the pressure by varying $\Phi$ with respect to the volume, while keeping temperature and chemical potential constant  
\be 
P = -\left( \frac{\partial \Phi}{\partial V} \right)_{T,\mu}. \label{pressure}
\ee

At $T=0$, we obtain the expression for $P$ using eqn.\eqref{lnz} in  \eqref{phitoz},
\be 
P = \int_{m_ec^2}^{\tilde{\e}_F} d\tilde{\e} \D (\tilde{\e}_F-\tilde{\e}), \label{kpressure} 
\ee
where we have replaced the summation with integral over energy.
Here, we have used the fact that at $T=0$, all the energy levels with energy greater than $E_F$ will be unoccupied.
We then obtain the relation,
\be 
P = n\tilde{\e}_F - \rho, \label{kprho}
\ee
with $n$ defined by
\be 
n = \int_{m_e c^2}^{\tilde{\e}_F} d\tilde{\e} \D = \int_{m_e c^2}^{\tilde{\e}_F} d\tilde{\e} \frac{4\pi g \vec{p}}{(2\pi k)^3} \frac{\tilde{\e}}{c^{2}} (1\mp \frac{a\tilde{\e}}{ck}). \label{number}
\ee
Note that the average energy $U$ is 
\be 
U = -\frac{\partial ln Z}{\partial \beta}.
\ee
and using \eqref{lnz} by replacing summation with integral, we find the energy density as given by
\be 
\rho = \frac{U}{V} = \int_{0}^{\infty} d\tilde{\e} \D \frac{\tilde{\e}}{e^{\beta(\tilde{\e}-\mu)}+1}.
\ee
This reduces for the case $T=0$ to
\be 
\rho = \int_{m_e c^2}^{\e_F} d\tilde{\e} \D \tilde{\e}. \label{krho}
\ee

Once we have the expression for the energy density, we could easily find out the expression for pressure. We first calculate the number density using the eqn.\eqref{number} for an electron in non-commutative spacetime, 
\bea 
n &=& \int_{m_e c^2}^{\e_F} \frac{1 }{\pi^2 k^3 c^{2}} p\tilde{\e} (1\mp \frac{a\tilde{\e}}{ck}) d\tilde{\e}, \\
&=&  \int_{0}^{p_F} \frac{1 }{\pi^2 k^3} p^{2} dp \mp  \int_{0}^{p_F} \frac{1 }{\pi^2 k^3} \frac{a}{k} p^{2}  \sqrt{p^{2}+ m_e c^{2}}  dp \label{cnumber}
\eea  
We find,
\bea 
n  = \frac{(m_e c)^{3}}{\pi^2 k^3} \frac{x_F^{3}}{3} \mp  \frac{(m_e c)^{4}}{\pi^2 k^3} \frac{a}{k} \frac{1}{8} \chi(x), \label{ncnumber}
\eea
where we have introduced a new variable, $x_F = \frac{p_F}{mc}$. The function $\chi(x)$, which is introduced to write the expression for number density in a compact form, has the form
\bea 
\chi(x) &=& \left( x(1+2x^2)\sqrt{1+x^2} - \ln(x+\sqrt{1+x^2})\right).
\eea

Demanding that number density to be always non-negative implies that  $ax \sim 10^{-22}$, when $x>>1$. This sets an upper bound to the deformation parameter as $10^{-22}$m. More on this will be discussed later.

We calculate the energy density $\rho$, using eqn.\eqref{ncdensity2} in \eqref{krho} and in the relativistic limit we get
\bea 
\rho &=&\int_{m_e c^2}^{\tilde{\e}_F} \frac{1 }{\pi^2 k^3 c^{2}} p \tilde{\e} (1\mp\frac{a}{k c}\tilde{\e}) \tilde{\e} d\tilde{\e}, \\
&=& \int_{m_e c^2}^{\tilde{\e}_F} \frac{1 }{\pi^2 k^3 }  p^2 \sqrt{(pc)^2+(m c^2)^2} dp \mp \int_0^{p_F}  \frac{1 }{\pi^2 k^3} \frac{a }{k c} p^{2}\sqrt{(pc)^2+(m c^2)^2}dp, \label{rint}
\eea
where we have used eqn.\eqref{tilep}. Eqn.\eqref{rint}, can be re-expressed as
\be  
\rho = \frac{(m_e c)^{4}c}{8 \pi^2 k^3} \chi(x) \mp \frac{(m_e c)^{5}}{\pi^2 k^3} \frac{ac}{k} \left(\frac{x^{5}}{5}+\frac{x^{3}}{3}\right).
\ee
 The $\mp$ sign in the equation is due to the fact that we have two different choices of $\varphi$, with each sign will corresponding to a specific choice of realization. Further, it is to be noted that in the commutative limit ($a\rightarrow 0$) this reduces to the familiar expression for density.
 
From, eqn.\eqref{kprho}, we have
 
%\ba 
%P &=  n\e_F - \rho = \frac{8 \pi}{(2\pi k)^3} \frac{p^{3}}{3} \sqrt{p_F^{2}+m^{2}} -   \\ &  \frac{8 \pi}{(2\pi k)^3} \frac{1}{8} \left[ \left(p_F \sqrt{m^2+p_F^2} \left(m^2+2 p_F^2\right) -m^4 \log \left( \sqrt{m^2+p_F^2}+p_F\right)\right) \right] + m^4 \log m \right). 
%\ea
\bea 
P = \frac{(m_e c)^{4}c}{\pi^2 k^3} \sqrt{x_F^2 +1}\frac{x_F^{3}}{3} \mp \frac{(m_e c)^{5}}{\pi^2 k^3} \frac{ac}{k} \frac{1}{8} \sqrt{x_F^2 +1}\chi(x_F) \nonumber \\
-\frac{(m_e c)^{4}c}{8 \pi^2 k^3} \chi(x) \pm \frac{(m_e c)^{5}}{\pi^2 k^3} \frac{ac}{k} \left(\frac{x^{5}}{5}+\frac{x^{3}}{3}\right). \label{presure}
\eea

In a compact star, the pressure is dominated by the pressure of the gas of Fermions. We assume that the temperature of the star is small, so that Fermi gas at $T=0$ is a good approximation (see for example, \cite{huang} for a nice discussion). Since energy density of white dwarf is dominated by nucleons, we write, to a good approximation
\be 
\rho = n_p m_p c^{2}+ n_e m_e c^{2}+ n_n m_n c^{2} \approx \alpha n_p m_p c^{2},
\ee
where suffix $p$ refers to proton, $n_p$ is number density of protons and the factor $\alpha$ takes into account the mean number density. Now, since the star is electrically neutral, we have $n_e = n_p$. Therefore, using eqn.\eqref{ncnumber}, we write
\be 
\rho = \alpha m_p c^{2} n_e = \alpha m_p c^{2} \frac{(m_e c)^{3}}{\pi^2 k^3} \frac{x_F^{3}}{3}.  \label{33}
\ee
We have neglected the correction to number density due to non-commutativity, as the correction term is negligible at the desired energy range.  Although, we do have a non-commutative correction for \eqref{33}, in writing the \eqref{33} we have neglected this term as this will only generate higher order corrections in $a$ when used in the expression for pressure given by \eqref{presure}.

Since in this letter, we are concerned only with very high energy effects, we will work with electrons at ultra-relativistic regime. By relativistic, we mean that the Fermi momentum of the electron to be much larger than it's rest mass, i.e., $x_F \gg 1$. In this limit, the function $\chi(x)$ introduced in eqn.\eqref{ncnumber} is
\be 
\chi(x) = 2 x^4. \label{chilimit}
\ee

Using $x_F = \frac{k}{m_e c} (3\pi^2 n_e)^{1/3} = \frac{k}{m_e c} (\frac{9\pi M}{4\mu m_p})^{1/3}\frac{1}{R}$ and using eqn.\eqref{chilimit} in \eqref{presure} to obtain 

\bea 
P = \frac{c}{\pi^2} \left( \frac{1}{12} k  \left(\frac{9\pi M}{4\mu m_p}\right)^{4/3}\frac{1}{R^4}  \mp \frac{1}{20} a c k  \left(\frac{9\pi M}{4\mu m_p}\right)^{5/3}\frac{1}{R^5}\right). \label{prho}
\eea

In the limit, $a\rightarrow 0$, we retrieve the commutative result. Also note that the $a$-dependent corrections in the above equation is independent of temperature as in the commutative case.

Backed with this result, we now turn to the study of stability of star under gravitational force. For a star to be gravitationally stable, we require that the change in its radius (assuming only radial force are present) due to gravitational force to balance the pressure due to the gas. Thus,
\be 
P \delta V = 4\pi R^{2} \delta R = f \frac{GM^{2}}{R^{2}} \delta R,
\ee
where $f$ is a number of order unity and depends on the functional form of density as function of radius $R$. We obtain the pressure as
\footnote{Gravitational constant $G$ can have corrections due to the effects of non-commutativity. For eg. in \cite{Gupta:2013ata}, it was 
shown, in the study of BTZ black hole, that $\frac{1}{G} \rightarrow \frac{1}{G} \left(1+ A \frac{a}{l^{2}} \sqrt{\frac{GM}{c^{2}}}\right)$, 
where $a$ is the deformation parameter and $A$ is a numerical constant. However, in here we are not considering such corrections of $G$ as it will
not change the bounds on $a$ considerably.}
\be 
P  = \frac{f}{4\pi} \frac{GM^{2}}{R^{4}}. \label{press}
\ee 
%Here, it is to be noted that the effect of $\kappa$-deformation will also led to a modification in the mass as studied in \cite{Harikumar:2010aw,Gupta:2017xex}.
%However, for the present analysis the correction due to mass modification in the expression for hydrostatic equillibrium is insignificant and thus we will not be keeping the mass modification in our equations.

 For the remaining analysis, we keep only terms of the order of $a$. This means that we write the expression for hydrostatic equillibrium as
\bea 
\frac{f}{4\pi} \frac{GM^{2}}{R^{4}} = \frac{\left(m_e c\right)^{4}c}{\pi^2 \hbar^3} \left( \frac{1}{12}\left(\frac{\hbar}{m_e c}\right)^4   \left(\frac{9\pi M}{4\mu m_p}\right)^{4/3}\frac{1}{R^4}  +\frac{1}{20}\frac{am_ec}{\hbar} \left(\frac{\hbar}{m_e c}\right)^5  \left(\frac{9\pi M}{4\mu m_p}\right)^{5/3}\frac{1}{R^5}\right). \label{equilli}
\eea
It should be emphasized that the above expression depends on two parameters, $f$ and $\mu$ which depend on the density distribution of matter inside star and its composition, respectively. Since, with only `$+$' sign we obtain a consistent physical expression, we choose the realization $\varphi = e^{+ap_0}$ for the rest of this paper. If we had chosen the realization, $\varphi = e^{-ap_0}$, we obtain a hydrostatic equilibrium condition which do not have a valid physical solution (see discussion after eqn.\eqref{p2}).

 Recall that the mass of the sun, which is a standard unit for measuring stellar masses is known upto first three decimal places exactly. To be precise, the current accepted value of solar mass is given by $M_{\odot} = 1.98848(9)\times 10^{30} kg$ {\cite{pdg}}. This means that deviation due to non-commutative term from commutative results is experimentally detectable, at the present level, if the ratio of second term in RHS of \eqref{equilli} to the first term is of the order of $10^{-3}$. In other words, we find that the effect of non-commutativity is significant when,
 \be 
\frac{12}{20} \left(\frac{9\pi}{8m_p}\right)^{\frac{1}{3}} \frac{a M^{\frac{1}{3}}}{R} \leq 10^{-3},
 \ee
or, plugging the value of proton mass into this expression, we have
\be 
\frac{a M^{\frac{1}{3}}}{R} \leq 1.298 \times 10^{-12}. \label{bound1}
\ee
Since, white dwarfs generically have mass of the order of solar mass, $M_{\odot}$ (i.e., $\sim 10^{30}$kg) and radius of the order $10^6$m. This implies a bound on the value of deformation parameter $a$ to be $10^{-16}m$. That is, the deformation parameter should be at the most as large as $10^{-16}m$ ($a<10^{-16}m$).   

Further, note that when contribution of the non-commutative term becomes much larger than the $a$-independent term, the expression obtained by neglecting the commutative term in eqn.\eqref{equilli}, is of the form
\be 
\frac{f}{4\pi} \frac{GM^{2}}{R^{4}}  + \frac{1}{20}\frac{am_ec}{\hbar} \left(\frac{\hbar}{m_e c}\right)^5  \left(\frac{9\pi M}{4\mu m_p}\right)^{5/3}\frac{1}{R^5} =0. 
\ee
The above equation do not have a physically valid solution and hence we will not consider this situation as it will lead to a solution of a star which is gravitationally unstable.

The above requirements yields us an upper bound on $a$ of the order of $10^{-10}m$. In other words, if the deformation parameter is larger than $10^{-10}m$ (i.e., $a>10^{-10}m$), then we do not have a physically acceptable white dwarf solution. This puts up another physical constraint on the value of deformation parameter. We have yet another constraint from \eqref{number} for the case $x\gg 1$. This is obtained from the physical condition that number density cannot be negative. Here we obtain that the deformation parameter should be atleast of the order of $10^{-22}$m. Thus we see that the most stringent condition we get on the deformation parameter is $a< 10^{-22}m$.

%Finally, suppose that $y$ be the ratio of second term to first term in RHS of \eqref{equilli}, then
%\be 
%\frac{a M^{\frac{1}{3}}}{m_p^{\frac{1}{3}} R} \sim y.
%\ee 
%Substituting the values $a = 10^{-22}m$, proton mass, $m_p \sim 10^{-27}kg$, $R \sim 10^{6}m$ and $M \sim 10^{30}$kg into the above expression, we find that $y \sim 10^{-9}$. This means that if the value of deformation parameter, $a\sim 10^{-22}m$, then for a typical white dwarf the deviation due non-commutativity is at the decimal places, which is negligible. However, 

\section{Alternative derivation for pressure using generalized uncertainty principle}
One of the consequence of quantum gravity is that the spacetime loses its operational meaning at energy scale compared to Planck scale. An interesting aspect of this effect is modification of uncertainty principle \cite{Doplicher:1994tu}.
Here in this section, we use generalized uncertainty principle to derive the expression for pressure inside a compact star. Let us consider a Fermi gas with uncertainty in the position of the particle given by $\Delta x$. For a Fermion restricted to region $\Delta x$, the pressure exerted is 

\be 
P = \frac{\text{Work done}}{volume}.
\ee
The work done, by the relativistic gas, on a region of size $\Delta x$ is $W = 2 \times (\Delta p) c$. The factor $2$ is because of the fact we consider a degenerate Fermi gas which can hold at most two Fermions per state. We thus write the expression of pressure for a gas of relativistic particle as (for a standard derivation in commutative spacetime, see for eg.\cite{huang})

\be 
P \approx \frac{2 c (\Delta p) }{(\Delta x)^{3}}. \label{pressur}
\ee
We now look for the effect of generalised uncertainty principle on the pressure. We use a generalised uncertainty relation consistent with $\kappa$-deformed theories, as shown in \cite{Anjana:2017thy}, i.e.,
\be 
(\Delta x) (\Delta p) (1+  \frac{\frac{3}{4} a}{\Delta x}) \ge \frac{\hbar}{2}. \label{gup1}
\ee 
 Note that in the limit $a \rightarrow 0$, this reduces to usual uncertainty principle.

The order of magnitude of this parameter is related to the length scale at which quantum gravity becomes important. The minimum possible volume occupied by $N$ Fermions, in accordance with Pauli exclusion principle is
\be 
V = \frac{N}{2} (\Delta x)^{3}. \label{vol}
\ee
Using eq.\eqref{gup1} and \eqref{vol} in \eqref{pressur}, we obtain
\be 
P \approx \frac{\hbar c}{2^{\frac{4}{3}}} \left(\frac{N}{V}\right)^{\frac{4}{3}} +  \frac{\hbar 3 a ~c}{ 4 \cdot 2^{\frac{5}{3}}} \left(\frac{N}{V}\right)^{\frac{5}{3}}. \label{p1}
\ee
This reduces to the well known results in commutative space when $a \rightarrow 0$.
The number density of Fermions inside a compact star of volume, $V$ is
\be 
n = \frac{N}{V} \sim \frac{MA}{Z m_p R^{3}}, \label{ndensity}
\ee
where $M$ is the total mass of the star, $Z$ is the atomic number of nuclei with which the star is composed of, $A$ is the total number of nuclei, $m_p$ is the mass of proton and $R$ is the radius of the star. We re-express \eqref{p1} as
\be 
P \approx \frac{\hbar c}{2^{\frac{4}{3}}} \left(\frac{Z}{A m_p}\right)^{\frac{4}{3}} \frac{M^{\frac{4}{3}}}{R^{4}} +  \frac{ \hbar 3 a~ c}{4 \cdot 2^{\frac{5}{3}}}  \left(\frac{Z}{A m_p}\right)^{\frac{5}{3}} \frac{M^{\frac{5}{3}}}{R^{5}}. \label{p2}
\ee Comparing the above expression with eqn.\eqref{prho}, we see that eqn.\eqref{p1} and \eqref{prho} are identical, modulo numerical factors, if we choose the realization, $\varphi = e^{+ap_0}$. Note that here we have kept terms upto first order in deformation parameter $a$.  Applying the analysis, analogous to the one applied in deriving eqn.\eqref{p2} from eqn.\eqref{equilli}, we find that the deformation parameter satisfies the bound, $a< 10^{-16}m$ for a typical white dwarf. Thus the GUP valid for $\kappa$-deformed spacetime  leads to the same upper bound in $a$ we have obtained from \eqref{equilli}.

\section{Conclusions and discussions}
In this letter, we have analyzed the stability of compact stars in the $\kappa$-deformed spacetime and have obtained upper bounds on the deformation parameter  $a$. We have treated the compact star as a degenerate Fermi gas, where degenerate pressure and gravitational pressure are in equilibrium. We have studied this problem in two different ways. First, we used well known tools of statistical mechanics to calculate degeneracy pressure. Here the effect of non-commutativity of the spacetime enters through the modified density of states. This modification is due to the change in energy momentum dispersion relation in $\kappa$-spacetime. This also leads to a modification of number density. In the second approach, the effect of $\kappa$-deformation is introduced through the GUP associated with the $\kappa$-spacetime. Scale dependent modification of Heisenberg uncertainty relation is characteristic feature of many approaches to quantum gravity. Here, we start with a generic form of GUP, valid for $\kappa$-spacetime which is parameterised by deformation parameter and using it evaluates the degeneracy pressure of Fermi gas. We then analyze the degeneracy pressure of Fermions in $\kappa$-spacetime and obtain a bound on the deformation parameter. This bound is exactly the same as the one obtained from the pressure evaluated using the partition function.

Summing up information from all the bounds, we conclude that to yield physically sensible result the deformation parameter must be lower than $10^{-22}$m (i.e., $a< 10^{-22}m$).

In doing this calculation we have assumed that $x_F =\frac{p_F}{m_e c}= \frac{k}{m_e c}(3\pi^2 n_e)^\frac{1}{3} \gg 1$. Taking order of magnitude estimate for an electron, we see that number density of electrons, $n_e \gg 10^{11} kg/m^3$. An order of magnitude estimate for the expression $x_F =\frac{p_F}{m_e c}= \frac{k}{m_e c}\left(\frac{M}{m_p} \right)^{\frac{1}{3}} \frac{1}{R}$ yields us the relation, $M^{\frac{1}{3}} \gg 10^2 R$ or $M \gg (10^6 kg/m^{3}) R^3$, where $M$ is the mass of star, R is the radius and $m_p$ is the proton mass. This means that for mass of order of solar mass the radius of the star for which our approximation is valid is roughly $10^5$ km. These estimates gives us a fair understanding of scale of stars for which our calculations are applicable. In addition, from the analysis of eqns. \eqref{equilli},\eqref{p2} along with eqn.\eqref{ncnumber}, we find that the magnitude of deformation parameter should be always less than $10^{-22}m$ ($a< 10^{-22}m$) for physically consistent results.

Taking into account of the physical parameters, such as mass of electron, radius of sun, solar mass and speed of light, and from eqn.\eqref{equilli} and eqn.\eqref{number} we find that the deformation parameter has an upper bound of the order of $10^{-22}$m. If the value of $a$ is larger than this bound (i.e., $a> 10^{-22}m$), we find that there are effects which are unphysical such as number density becoming negative. Since, clearly, this is contrary to what are being observed in nature, this bound on $a$ is a strong physical constraint on value of deformation parameter,$a$. 

In addition to the derivation of the constraints on the deformation parameter $a$, we are able to fix the choice of realization of $\kappa$-spacetime by utilizing the generalised uncertainty relation obtained in \cite{Anjana:2017thy}. If we compare the expression of the pressure given in eqn.\eqref{prho} for degenerate Fermi gas using statistical mechanics and the pressure obtained by utilizing GUP as given in eqn.\eqref{p2}, we find that the realization should be $\varphi = e^{+ap_0}$, for both the expressions to be equivalent. This helps us in fixing the sign in the choice of $\varphi$. This is crucial in the sense that once we identify the realization that describe nature in the correct manner, we could use this realization in the future investigations to understand the consequences of noncommutativity in different physical systems. 

\section*{Acknowledgement}
ZNS would like to acknowledge the support for this work received from CSIR, Govt. of India under CSIR-SRF scheme.
EH thanks SERB, Govt. of India, for support through EMR/2015/000622.
 
\bibliographystyle{utphys} % Tell bibtex which bibliography style to use
\bibliography{ref.bib}

\providecommand{\href}[2]{#2}\begingroup\raggedright\begin{thebibliography}{10}

\bibitem{Doplicher:1994tu}
S.~Doplicher, K.~Fredenhagen, and J.~E. Roberts, ``{The Quantum structure of
  space-time at the Planck scale and quantum fields},'' {\em Commun. Math.
  Phys.} {\bf 172} (1995) 187--220,
  \href{http://xxx.lanl.gov/abs/hep-th/0303037}{{\tt hep-th/0303037}}.

\bibitem{Khan:2007fc}
S.~Khan, B.~Chakraborty, and F.~G. Scholtz, ``{On the role of twisted
  statistics in the noncommutative degenerate electron gas},'' {\em Phys. Rev.}
  {\bf D78} (2008) 025024, \href{http://xxx.lanl.gov/abs/0707.4410}{{\tt
  0707.4410}}.

\bibitem{KowalskiGlikman:2001ct}
J.~Kowalski-Glikman, ``{Doubly special quantum and statistical mechanics from
  quantum kappa Poincare algebra},'' {\em Phys. Lett.} {\bf A299} (2002)
  454--460, \href{http://xxx.lanl.gov/abs/hep-th/0111110}{{\tt
  hep-th/0111110}}.

\bibitem{AmelinoCamelia:2009tv}
G.~Amelino-Camelia, N.~Loret, G.~Mandanici, and F.~Mercati, ``{UV and IR
  quantum-spacetime effects for the Chandrasekhar model},'' {\em Int. J. Mod.
  Phys.} {\bf D21} (2012) 1250052,
  \href{http://xxx.lanl.gov/abs/0906.2016}{{\tt 0906.2016}}.

\bibitem{Shariati:2010zz}
A.~Shariati, M.~Khorrami, and A.~H. Fatollahi, ``{Statistical mechanics of free
  particles on space with Lie-type noncommutativity},'' {\em J. Phys.} {\bf
  A43} (2010) 285001, \href{http://xxx.lanl.gov/abs/1104.1486}{{\tt
  1104.1486}}.

\bibitem{Scholtz:2012tx}
F.~G. Scholtz and B.~Chakraborty, ``{Spectral triplets, statistical mechanics
  and emergent geometry in non-commutative quantum mechanics},'' {\em J. Phys.}
  {\bf A46} (2013) 085204, \href{http://xxx.lanl.gov/abs/1206.5119}{{\tt
  1206.5119}}.

\bibitem{Hosseinzadeh:2015jra}
V.~Hosseinzadeh, M.~A. Gorji, K.~Nozari, and B.~Vakili, ``{Noncommutative
  spaces and covariant formulation of statistical mechanics},'' {\em Phys.
  Rev.} {\bf D92} (2015), no.~2 025008,
  \href{http://xxx.lanl.gov/abs/1506.04164}{{\tt 1506.04164}}.

\bibitem{AmelinoCamelia:1999pm}
G.~Amelino-Camelia and S.~Majid, ``{Waves on noncommutative space-time and
  gamma-ray bursts},'' {\em Int. J. Mod. Phys.} {\bf A15} (2000) 4301,
  \href{http://xxx.lanl.gov/abs/hep-th/9907110}{{\tt hep-th/9907110}}.

\bibitem{AmelinoCamelia:2008qg}
G.~Amelino-Camelia, ``{Quantum-Spacetime Phenomenology},'' {\em Living Rev.
  Rel.} {\bf 16} (2013) 5, \href{http://xxx.lanl.gov/abs/0806.0339}{{\tt
  0806.0339}}.

\bibitem{Borowiec:2009ty}
A.~Borowiec, K.~S. Gupta, S.~Meljanac, and A.~Pachol, ``{Constarints on the
  quantum gravity scale from kappa - Minkowski spacetime},'' {\em Europhys.
  Lett.} {\bf 92} (2010) 20006, \href{http://xxx.lanl.gov/abs/0912.3299}{{\tt
  0912.3299}}.

\bibitem{AmelinoCamelia:2001vz}
G.~Amelino-Camelia, ``{Observed threshold anomalies as the first hope of a
  manifestation of Planck length physics},'' in {\em {Recent developments in
  theoretical and experimental general relativity, gravitation and relativistic
  field theories. Proceedings, 9th Marcel Grossmann Meeting, MG'9, Rome, Italy,
  July 2-8, 2000. Pts. A-C}}, pp.~1485--1492, 2001.
\newblock \href{http://xxx.lanl.gov/abs/gr-qc/0106005}{{\tt gr-qc/0106005}}.

\bibitem{Li:2009mt}
M.~Li, Y.~Pang, and Y.~Wang, ``{Non-Commutativity, Teleology and GRB Time
  Delay},'' {\em Phys. Lett.} {\bf B682} (2010) 334--336,
  \href{http://xxx.lanl.gov/abs/0904.1079}{{\tt 0904.1079}}.

\bibitem{Srivastava:2012av}
R.~Srivastava, ``{Signatures of New Physics from HBT Correlations in UHECRs},''
  {\em Mod. Phys. Lett.} {\bf A27} (2012), no.~28 1250160,
  \href{http://xxx.lanl.gov/abs/1201.2380}{{\tt 1201.2380}}.

\bibitem{Amati:1988tn}
D.~Amati, M.~Ciafaloni, and G.~Veneziano, ``{Can Space-Time Be Probed Below the
  String Size?},'' {\em Phys. Lett.} {\bf B216} (1989) 41--47.

\bibitem{Garay:1994en}
L.~J. Garay, ``{Quantum gravity and minimum length},'' {\em Int. J. Mod. Phys.}
  {\bf A10} (1995) 145--166, \href{http://xxx.lanl.gov/abs/gr-qc/9403008}{{\tt
  gr-qc/9403008}}.

\bibitem{Kempf:1996nk}
A.~Kempf and G.~Mangano, ``{Minimal length uncertainty relation and ultraviolet
  regularization},'' {\em Phys. Rev.} {\bf D55} (1997) 7909--7920,
  \href{http://xxx.lanl.gov/abs/hep-th/9612084}{{\tt hep-th/9612084}}.

\bibitem{AmelinoCamelia:1997jx}
G.~Amelino-Camelia, J.~Lukierski, and A.~Nowicki, ``{kappa deformed covariant
  phase space and quantum gravity uncertainty relations},'' {\em Phys. Atom.
  Nucl.} {\bf 61} (1998) 1811--1815,
  \href{http://xxx.lanl.gov/abs/hep-th/9706031}{{\tt hep-th/9706031}}. [Yad.
  Fiz.61,1925(1998)].

\bibitem{Maggiore:1993kv}
M.~Maggiore, ``{The Algebraic structure of the generalized uncertainty
  principle},'' {\em Phys. Lett.} {\bf B319} (1993) 83--86,
  \href{http://xxx.lanl.gov/abs/hep-th/9309034}{{\tt hep-th/9309034}}.

\bibitem{Capozziello:1999wx}
S.~Capozziello, G.~Lambiase, and G.~Scarpetta, ``{Generalized uncertainty
  principle from quantum geometry},'' {\em Int. J. Theor. Phys.} {\bf 39}
  (2000) 15--22, \href{http://xxx.lanl.gov/abs/gr-qc/9910017}{{\tt
  gr-qc/9910017}}.

\bibitem{Das:2008kaa}
S.~Das and E.~C. Vagenas, ``{Universality of Quantum Gravity Corrections},''
  {\em Phys. Rev. Lett.} {\bf 101} (2008) 221301,
  \href{http://xxx.lanl.gov/abs/0810.5333}{{\tt 0810.5333}}.

\bibitem{Nozari:2006gg}
K.~Nozari and B.~Fazlpour, ``{Generalized uncertainty principle, modified
  dispersion relations and early universe thermodynamics},'' {\em Gen. Rel.
  Grav.} {\bf 38} (2006) 1661--1679,
  \href{http://xxx.lanl.gov/abs/gr-qc/0601092}{{\tt gr-qc/0601092}}.

\bibitem{KalyanaRama:2001xd}
S.~Kalyana~Rama, ``{Some consequences of the generalized uncertainty principle:
  Statistical mechanical, cosmological, and varying speed of light},'' {\em
  Phys. Lett.} {\bf B519} (2001) 103--110,
  \href{http://xxx.lanl.gov/abs/hep-th/0107255}{{\tt hep-th/0107255}}.

\bibitem{Bemfica:2005pz}
F.~S. Bemfica and H.~O. Girotti, ``{The Noncommutative degenerate electron
  gas},'' {\em J. Phys.} {\bf A38} (2005) L539,
  \href{http://xxx.lanl.gov/abs/quant-ph/0506191}{{\tt quant-ph/0506191}}.

\bibitem{Meljanac:2006ui}
S.~Meljanac and M.~Stojic, ``{New realizations of Lie algebra kappa-deformed
  Euclidean space},'' {\em Eur. Phys. J.} {\bf C47} (2006) 531--539,
  \href{http://xxx.lanl.gov/abs/hep-th/0605133}{{\tt hep-th/0605133}}.

\bibitem{Chaichian:2004za}
M.~Chaichian, P.~P. Kulish, K.~Nishijima, and A.~Tureanu, ``{On a
  Lorentz-invariant interpretation of noncommutative space-time and its
  implications on noncommutative QFT},'' {\em Phys. Lett.} {\bf B604} (2004)
  98--102, \href{http://xxx.lanl.gov/abs/hep-th/0408069}{{\tt hep-th/0408069}}.

\bibitem{Balachandran:2006pi}
A.~P. Balachandran, T.~R. Govindarajan, G.~Mangano, A.~Pinzul, B.~A. Qureshi,
  and S.~Vaidya, ``{Statistics and UV-IR mixing with twisted Poincare
  invariance},'' {\em Phys. Rev.} {\bf D75} (2007) 045009,
  \href{http://xxx.lanl.gov/abs/hep-th/0608179}{{\tt hep-th/0608179}}.

\bibitem{Govindarajan:2008qa}
T.~R. Govindarajan, K.~S. Gupta, E.~Harikumar, S.~Meljanac, and D.~Meljanac,
  ``{Twisted statistics in kappa-Minkowski spacetime},'' {\em Phys. Rev.} {\bf
  D77} (2008) 105010, \href{http://xxx.lanl.gov/abs/0802.1576}{{\tt
  0802.1576}}.

\bibitem{huang}
K.~Huang, {\em Statistical mechanics, 2ed}.
\newblock Wiley India, 2008.

\bibitem{pal}
P.~B. Pal, {\em An introductory course of Statistical Mechanics}.
\newblock Narosa Publishing House, New Delhi, India, 2007.

\bibitem{Chandrasekhar:1931ih}
S.~Chandrasekhar, ``{The maximum mass of ideal white dwarfs},'' {\em Astrophys.
  J.} {\bf 74} (1931) 81--82.

\bibitem{Meljanac:2007xb}
S.~Meljanac, A.~Samsarov, M.~Stojic, and K.~S. Gupta, ``{Kappa-Minkowski
  space-time and the star product realizations},'' {\em Eur. Phys. J.} {\bf
  C53} (2008) 295--309, \href{http://xxx.lanl.gov/abs/0705.2471}{{\tt
  0705.2471}}.

\bibitem{Gupta:2013ata}
K.~S. Gupta, E.~Harikumar, T.~Juric, S.~Meljanac, and A.~Samsarov, ``{Effects
  of Noncommutativity on the Black Hole Entropy},'' {\em Adv. High Energy
  Phys.} {\bf 2014} (2014) 139172,
  \href{http://xxx.lanl.gov/abs/1312.5100}{{\tt 1312.5100}}.

\bibitem{pdg}
{\bf Particle Data Group} Collaboration, K.~A. Olive {\em et.~al.}, ``{Review
  of Particle Physics},'' {\em Chin. Phys.} {\bf C38} (2014) 090001.

\bibitem{Anjana:2017thy}
V.~Anjana, E.~Harikumar, and A.~K. Kapoor, ``{Non-Commutative space-time and
  Hausdorff dimension},'' \href{http://xxx.lanl.gov/abs/1704.07105}{{\tt
  1704.07105}}.

\end{thebibliography}\endgroup

\end{document}